\documentclass{aa}
\usepackage{natbib}
\bibpunct{(}{)}{;}{a}{}{,}  
\usepackage{graphicx}
\usepackage{txfonts}
\usepackage{aalongtable}
\begin{document}

\title{An X-ray Survey in SA 57 with XMM-Newton\thanks{Based on observations obtained with XMM-Newton, an ESA science mission with instruments and contributions directly funded by ESA Member States and NASA}}

%   \subtitle{}

\author{D. Trevese \inst{1}, F. Vagnetti \inst{2}, S. Puccetti \inst{3,4}, F. Fiore \inst{4}, 
M. Tomei \inst{1}, M. A. Bershady \inst{5}}

   \offprints{Dario Trevese \email{dario.trevese@roma1.infn.it}}

   \institute{Dipartimento di Fisica, Universit\'a di Roma ``La Sapienza'', P.le A. Moro 2, 00185 Roma (Italy)            
         \and
	 Dipartimento di Fisica, Universit\'a di Roma ``Tor Vergata'', Via delle Ricerca Scientifica 1, 00133 Roma (Italy) 
         \and
	 ASI Science Data Centre, c/o ESRIN, via G. Galilei, I-00044 Frascati (Italy)
         \and
	 INAF - Osservatorio Astronomico di Roma, Via di Frascati 33, 00040 Monte Porzio Catone (Italy)
         \and
	 Department of Astronomy,  University of Wisconsin, 475 North Charter Street, Madison, WI 53706 (U.S.A.)
             }

 \date{}

  \abstract
{The maximum number density of Active Galactic Nuclei (AGNs), as deduced from X-ray studies,
occurs at $z\la$1, with lower luminosity objects peaking at smaller
redshifts. Optical studies lead to a different evolutionary behaviour,
with a number density peaking at $z\approx$2 independently of the
intrinsic luminosity, but this result is limited to active nuclei
brighter than the host galaxy.  A selection based on optical
variability can detect low luminosity AGNs (LLAGNs), where the host
galaxy light prevents the identification by non-stellar colours.}
{We want to collect X-ray data in a field where it exists an
optically-selected sample of ``variable galaxies'', i.e. variable
objects with diffuse appearance, to investigate the X-ray and optical
properties of the population of AGNs, particularly of low luminosity
ones, where the host galaxy is visible.}
{We observed a field of ~0.2 deg$^2$ in the Selected Area 57, for 67
ks with XMM-Newton. We detected X-ray sources, and we correlated the
list with a photographic survey of SA 57, complete to B$_J \sim 23$
and with available spectroscopic data.}
{We obtained a catalogue of 140 X-ray sources to limiting fluxes $5
\times 10^{-16}$, $2 \times 10^{-15}$ erg cm$^{-2}$ s$^{-1}$ in the
0.5-2 keV and 2-10 keV respectively, 98 of which are identified in the
optical bands. The X-ray detection of part of the variability-selected
candidates confirms their AGN nature. Diffuse variable objects
populate the low luminosity side of the sample.  Only 25/44
optically-selected QSOs are detected in X-rays. 15\% of all QSOs in
the field have $X/O < 0.1$.}
{}

\authorrunning{D. Trevese et al.}
\titlerunning{SA57XMM survey}
\maketitle

\section {Introduction}

Supermassive black holes  (SMBHs) are believed to inhabit  most, if not
all,  bulges of  present-epoch galaxies \citep{kor95}, 
and  strong evidences exist of  a correlation
between  the  black hole  mass  and  either  the mass  and  luminosity
\citep[][and refs. therein]{mar03} or the  velocity dispersion of the
host  bulge \citep{fer00,tre02}.  This strongly
suggests  that  the formation  and  growth  of  SMBHs and galaxies  are
physically  related processes. A  theory of  cosmic structure formation
and the  nature  of the Active Galactic Nuclei (AGN)  feedback \citep{silk98,cav02,vitt05}
requires the knowledge of the evolution  in  cosmic time  of  the AGN  population.

In recent years consensus has grown on a fast increase of the number
density of QSOs moving forward in cosmic time, until $z\sim$3,
followed by a slower decline of the luminosity function (LF), which
can be described by a QSO luminosity evolution.  The quantification of
this behaviour is currently based on the 2QZ survey \citep{cro01} for
$z<$ 2.5, on \citet{war94} and \citet{schm95} surveys for $z\ge$ 3 and
on the Sloan Digital Sky Survey (SDSS) data for $z>$ 4.5
\citep{fan01,and01}.  None of the above surveys covers the redshift
region where the maximum of QSOs density is located. Moreover, higher
redshift data are restricted to the bright end of QSO LF and even low
redshift data do not sample the evolution of objects fainter than $M_B
< -23$.

\citet{wol03} analyse the
intermediate redshift region, where the  maximum in cosmic time of the number-density of AGNs is
located ($z\sim 2$).  They use a sample selected by a multi-band technique, and
extend the study down to $M_B \simeq -21.5$.  They provide the most
accurate measurement available to date of the maximum in cosmic time of
the AGN comoving space density.  However their method cannot
select fainter AGNs due to the contribution of the host galaxy light
to the observed spectral energy distribution.
Variability was adopted as a tool to select a  sample
of QSOs with point-like images in  SA 57, on the 
basis of a collection of 
photographic plates taken at the Mayall 4m KPNO telescope about once per year for
15 consecutive years \citep{t89}. This technique is well suited for selection of 
intrinsically low luminosity AGNs (LLAGNs), since variability is higher in AGNs of lower 
luminosity \citep{t94,hook94,cris96,vand04}. For these reasons a sample 
of "variable galaxies", i.e. variable objects with extended images,
was created from the same plates of SA 57 \citep[][(BTK)]{btk98}.
Spectroscopic observations have already confirmed the
AGN nature of 5 relatively bright ($B_J <$ 22.5) objects, and provided
redshifts in the range $0.2 < z < 0.4$ and absolute B magnitudes
$-22.5 < M_B < -19.0$. Subsequently, \citet{sara03} selected galaxies with variable nuclei in the Hubble Deep Field, showing that a sizable fraction of them is undetected in the X-rays even at the flux limits of the 2Ms Chandra Deep Field North Survey \citep{alex03}. Thus, optical variability is a good complementary AGN selection criterion, which is also competitive, with respect to X-ray surveys, to efficiently find high sky densities of AGNs \citep{bran05}.

Hard X-ray observations are the most efficient way to discriminate
between accretion-powered sources, such as AGN, from starlight and
optically thin, hot-plasma emission.  Deep surveys have resolved
80-90\% of the 2-10 keV Cosmic X-ray Background (CXB) into
sources \citep{more03,bran05,hick06}.  While these studies are, at
least qualitatively, confirming the predictions of standard AGN
synthesis models for the CXB \citep[e.g.][]{coma01}, somewhat
surprising results are also emerging: i) the sources making the CXB have a maximum number density at a redshift ($z\sim1$), lower than predicted by synthesis models
\citep[e.g.][]{hasi03}; ii) there is evidence of a strong luminosity
dependence to the evolution, with low luminosity sources (i.e.
Seyfert galaxies) peaking at a significantly later cosmic time than
high luminosity ones \citep{hasi03,cowi03,fior03,lafr05}.  However, a
direct comparison with the optical LF evolution requires to extend the
latter down to $M_B\sim -19$, i.e. in the range of the BTK sample.

For this reason we have performed a medium-deep observation of the same field in 
the SA 57 which is one of the best studied fields  of the sky at
all  wavelengths: radio  FIRST Survey  \citep{beck95},  IR deep
ISOPHOT  Survey \citep{lind00},  soft  X-ray ROSAT  HRI
\citep{miya97}.  A  field of $\sim  35$ arcmin in  diameter has
been repeatedly observed  since 1975 in the $U$, $B_J$, $F$, $N$ bands. 
A number of search techniques,  including non-stellar  colour, absence  of  proper  motion
and variability have been applied in this field
for the detection of QSOs/AGNs to faint limits \citep{kkc86,koo88,t89,t94,btk98}.

We observed the central area of SA 57 with XMM-Newton with the aim of
performing a combined X-ray and optical analysis of the sources
detected in the field.  As a result we produced a catalogue of 140
X-ray sources.  In the present paper we report on the results of these
observations and the optical identification made possible by the
already existing photometric and spectroscopic data. This allows a
break down of a substantial fraction of the X-ray sample into normal
galaxies, different types of Seyfert galaxies, QSOs and possibly
obscured {\it quasar-2} type objects.

The paper is organised as follows.
Section 2 describes the X-ray sample,
Section 3 discusses the known optical sources, both detected and undetected in X-rays,  
Section 4 discusses the results, 
Section 5 contains a summary.

Consensus cosmology, $H_o=75$ km s$^{-1}$Mpc$^{-1}$, $\Omega_m=0.3$,  $\Omega_{\Lambda}=0.7$ is adopted throughout the paper.

\section {X-ray sources}

A deep XMM-Newton pointing covers the SA57 region. The pointing was centred
at $\alpha=13^h08^m28^s$ and $\delta=29^o23'07''$ (J2000).  The X-ray
observations were performed on January 2005 with the European Photon Imaging
Camera (EPIC: one PN-CCD camera (0.5-10 keV, \citet{stru01}) and two MOS-CCD
cameras (MOS1, MOS2, 0.3-10 keV, \citet{turn01}). Table  \ref{Tab1} gives a log of the
XMM-Newton observations.

The data have been processed using the XMM-Newton Science Analysis
Survey (SAS) v.6.0. We used the event files linearised with a standard
reduction pipeline (Pipeline Processing System, PPS) at the Survey Science
Centre (SSC, University of Leicester, UK). Events spread at most in two
contiguous pixels for PN (i.e. pattern=0-4) and in four contiguous pixels for
MOS (i.e, pattern=0-12) have been selected.  Event files were cleaned from bad
pixels (hot pixels, events out of the field of view, etc.) and the soft proton
flares following \citet{pucc06}.

Source detection was performed on co-added (in sky coordinates)
PN+MOS1+MOS2 images accumulated in four energy bands: 0.5-10 keV (total
band, T), 0.5-2 keV (soft band, S), 2-10 keV (hard band, H), 5-10 keV
(ultra hard band, HH). 

\begin{table}
\begin{center}
\caption{ \label{Tab1} Observation log.}
\begin{tabular}{lcccc}
\hline
Instrument &  Exposure [ksec]& Net exposure [ksec]$^a$ \\
\hline
\hline
PN &  62   & 45  \\
MOS1 & 67  & 49     \\
MOS2  & 67   & 51     \\
\hline
\hline
\end{tabular}
\end{center}
$^a$Net on-axis exposure time after rejection of high background periods (see
Sect. 2).  
\end{table}

The source detections and the X-ray photometry were performed by using the
PWXDetect code, developed at INAF~-~Osservatorio Astronomico di Palermo,
following \citet{pill06}. The code is derived from the original
ROSAT code for source detection by \citep{dami97} and allows one to combine
data from different EPIC cameras and data taken in different observations, in
order to achieve the deepest sensitivity.  The code is based on the analysis
of the wavelet transform (WT) of the count rate image. A WT of a
two-dimensional image is a convolution of the image with a ``generating
wavelet'' kernel, which depends on position and length scale.  In the
algorithm developed by \citet{dami97}, the generating wavelet is a ``Mexican
hat''. The length scale is a free parameter; therefore this method is
particularly suited for cases where the point spread function (PSF) is
strongly varying across the image, and moreover includes the exposure maps to
handle sharp background gradients. It also provides robust detections of
extended sources.

To evaluate count rates, we have chosen PN as the reference
detector. The scaling factor  between PN, MOS1 and MOS2 depends on
the relative instrument efficiency and source spectral shape. We used the
scaling factors evaluated by Puccetti et al. (2006) for a power law model with an energy
index $\alpha_E=0.8$.  

We adopted a threshold on the significance level corresponding to a probability of $2 \times 10^{-5}$ that a 
local maximum is generated by a Poisson fluctuation of the background counts.
The limiting fluxes on axis, in the T, S, H and HH bands, are approximately  $10^{-15}$,  $5 \times 10^{-16}$, $2 \times 10^{-15}$, $10^{-14}$  erg cm$^{-2}$  s$^{-1}$ respectively.
The result of this selection is a sample of 140 sources, detected in at least one band.

For each source, count rates were converted to fluxes adopting constant conversion factors for the S and H bands 
$f_S=a_S c_S$,  $f_H=a_H c_H$, with $a_S=1.61 \times 10^{-12}$ erg cm$^{-2}$, $a_H=8.26 \times 10^{-12}$erg cm$^{-2}$,
corresponding to an ``average source spectrum'' $f_{\lambda}^{A}$ represented
by a power law of energy index $\alpha_E=0.8$.
Fluxes $f_T$ in the T band are computed as follows:
a) if the object is detected in 
both the  S and H bands, $f_T=f_S+f_H$;
b) if the object is detected  in the T band and  in  one of the S or H bands, 
the missing count rate is evaluated as $c_H=c_T-c_S$ or $c_S=c_T-c_H$ respectively, and the flux is computed as above
and reported in Table  \ref{Tab1} if greater than zero;
c) if the object is detected only in S or H band $f_T=f_S$ or $f_T=f_H$ respectively;
d) if the object is detected in the T band only, the flux  $f_T$ reported in the table
 is computed adopting an average conversion factor, $f_T=a_T c_T$ with $a_T=3.75 \times 10^{-12}$ erg cm$^{-2}$ 
 determined by the linear regression of $f_T$ vs. $c_T$ for all the objects of cases a), b), c).
In any of the bands, when the source is not detected and the flux cannot be evaluated from the other bands,
a 3-$\sigma$ upper limit is computed adopting the  relevant average conversion factor.

The results are reported in Table \ref{Tab2}, where 
{\it column 1}: catalogue serial number;
{\it columns 2,3}: right ascension and declination (J2000); 
{\it columns 4,5,6}: fluxes in the T, S, H bands respectively; 
{\it column 7}: the identification rank, as defined in the note. 
3-$\sigma$ upper limits are preceded by a ``$<$'' character.

\begin{figure}
   \centering
\resizebox{\hsize}{!}{\includegraphics{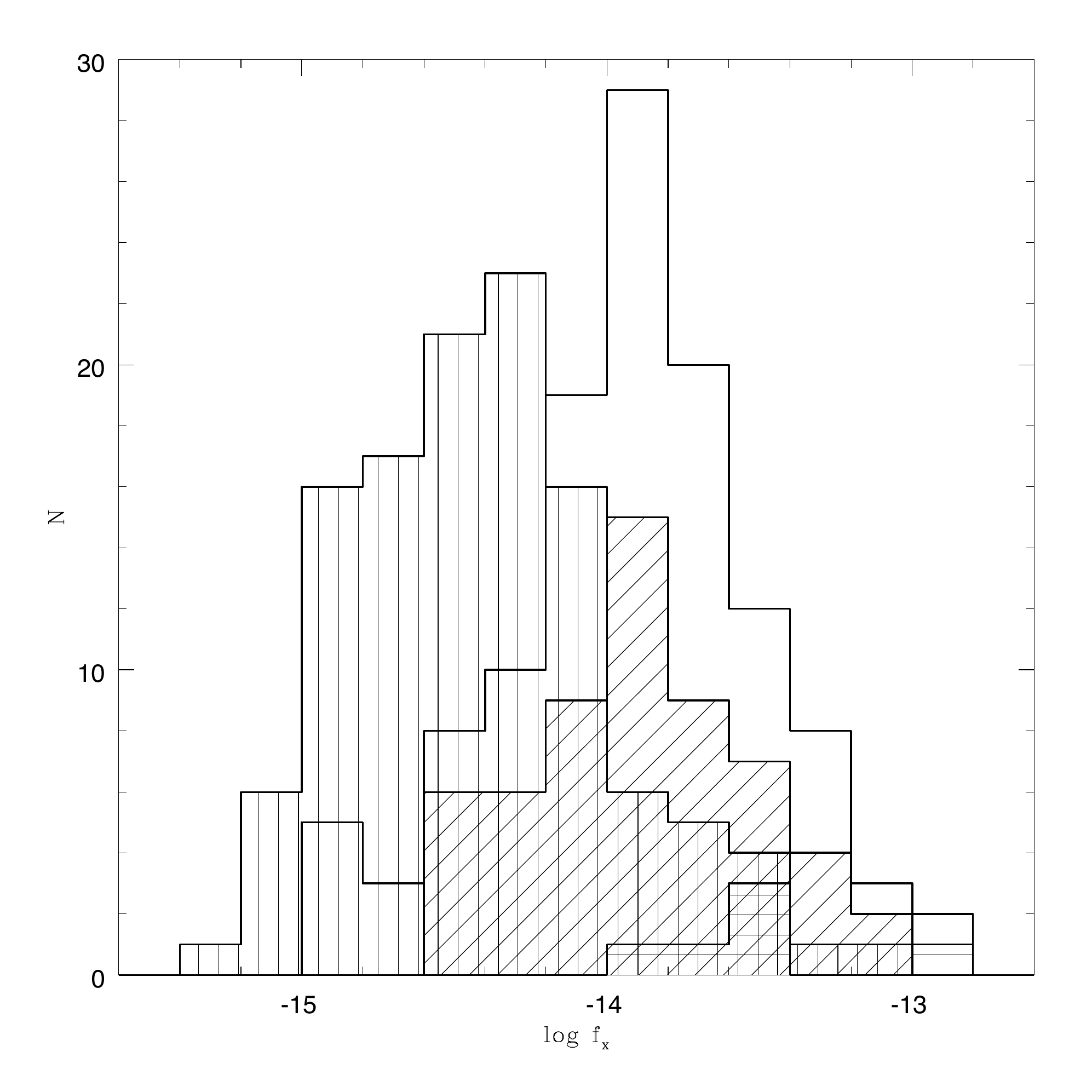}}

      \caption{
The flux distribution for the sources detected in the different bands: T band (no shading), S band (vertical shading), 
H band (diagonal shading), HH band (horizontal shading). 
}
         \label{Fig1}
   \end{figure}

Of the 140 sources, the numbers of sources detected in the
T, S, H an HH bands are 119, 117, 58, 6  respectively and
the relevant histograms are shown in Figure \ref{Fig1}. 

%%%%%%%%%%%%%% Figure 1 %%%%%%%%%%%%%%%%

%%%%%%%%%%%%%%%%%%%%%%%%%%%%%%%%%%%%%%%

On the basis of the optical coordinates of the  most secure   quasar identifications in the field, a small shift in 
$\alpha$ and $\delta$ has been computed, respect to the (X-ray) detection coordinates, to obtain a more accurate 
correspondence with our optical coordinate system, based on USNO-A2.0 (see next section).

\section{Optical sources}

A photographic  survey of  SA57 was conducted
with the prime  focus camera at Mayall 4 m telescope at  Kitt Peak  National Observatory (KPNO)
from 1974 to 1989. A photometric catalogue of   8146 objects in 
$U$, $B_J$, $F$ and $N$ bands in a  field of $\sim 0.3$
deg$^2$,   complete  to   $B_J   \sim  23$  was   used  for   optical
identifications  of  the  X-ray  detected sources \citep{kron80,koo86}. 
The optical coordinates were recomputed by cross-correlating the SA57 catalogue
with USNO-A2.0 catalogue. After a 2-$\sigma$ rejection, the IRAF {\it ccmap} utility
provides a 4th order coordinate transformation based on 446 objects spread over the field,
with $<$0.2 arcsec r.m.s. deviation in both $\alpha$ and $\delta$, respect to USNO-A2.0.

The above multi-epoch observations have been used to select AGN candidates on the basis of variability. In the case of point-like sources, the variability criterion has proven to be 74\% complete (variable AGNs/total AGNs), while its reliability  (variable AGNs/total variables) is between 80\% and 95\%, depending on the variability threshold adopted, which in turn is chosen taking into account the photometric accuracy \citep{t89}. For the extended objects, variability can select AGNs even in cases where color selection is not effective due to dominance of the host galaxy light. A sample of 51 variable extended objects was created by \citet{btk98} (BTK), 16 of which have $B_J<22.5$. A fraction of them was observed spectroscopically, and 5 objects confirmed their AGN character. A new spectroscopic campaign of SA57 is in progress \citep{trev07}. In the meantime X-ray emission from some of these candidates strongly suggests their AGN nature (see below).

\subsection{Optical identification of X-ray sources}

The optical identifications are reported in Table \ref{Tab3}, where the columns have the following meaning:
{\it column 1}: source serial number;
{\it column 2}: identification rank (same as Table \ref{Tab2}); 
{\it column 3}: serial number NSER in the optical catalogue of KPNO survey of SA57 \citep{kron80,koo86}, 
or, preceded by a ``G'', in the  NGPFG catalogue \citep{infa95}.
{\it columns 4,5}: right ascension and declination (J2000);  
{\it columns 6,7}:  $B_J$ and $F$  magnitudes respectively; 
{\it column 8}: redshift; 
{\it column 9}: source class (as specified in the note).

Identifications indicated with ``I'' are the most secure and
correspond to optical positions within 5 arcsec from the X-ray
position, and no other optical objects inside this area. Other
marginal or less secure identification, more distant than 5 arcsec,
are indicated with ``M''. When more than one object falls within 5
arcsec from the X-ray position sources are indicated with ``A'',
meaning ambiguous identification.  A total of 98 objects has been
identified with optical sources. Of these, 72 are most secure
identifications (I), 15 are classified as marginal (M), 11 are
ambiguous (A).  42 sources are unidentified (U).

Of the 72 most secure identifications 33 are either confirmed or
candidate AGNs: 24 confirmed and 3 candidate QSOs, 5 BTK objects (two
of which are spectroscopically confirmed), and 1 radio galaxy.
Another 7 optically identified X-ray sources correspond to galaxies
with spectroscopic redshifts in the catalogue of \citet{munn97} of SA
57. One of these corresponds to the cD galaxy in the
centre of the galaxy cluster II Zw 1305.4+2941 at $z=0.241$.  Among the AGN candidates selected through variability alone (i.e., not previously selected by other methods), 9 are detected securely in X-rays (4 point-like and 5 extended, see note ``e'' in
Table \ref{Tab3}).  Of these, 3 are already confirmed by optical
spectroscopy. For the remaining 6 objects we can provisionally assume the
X-ray detection as a confirmation of their AGN nature, though optical
spectroscopy will be eventually necessary to measure their absolute
luminosity and assign them to a specific class.

\subsection{X-ray undetected optical sources}
We have searched the NASA Extragalactic Database (NED) for objects with known redshift 
and classified as ``QSO'', in the area covered by our X-ray survey. 
The resulting list contains 44 objects, 25 of which already appear in Table \ref{Tab2} and Table \ref{Tab3}
since they are detected in X-rays. The remaining 19 QSOs not detected in X-rays are reported in Table \ref{Tab4}.
Most of them  belong to the \citet{kkc86} list, i.e., colour selected, point-like AGN candidates.
For all of them we computed the 3-$\sigma$ upper limits to  X-ray fluxes.
The  meaning of columns in Table \ref{Tab4} is the following:
{\it column 1}: serial number NSER in the optical catalogue of KPNO survey of SA57 \citep{kron80,koo86};
{\it column 2,3}: $\alpha,\delta$ (2000);
{\it column 4}: F; 
{\it column 5}: redshift;
{\it column 6}: 3-$\sigma$ upper limit in the 2-10 keV band. 

We have also considered 21 X-ray undetected  objects, which were selected as AGN candidates on the basis of their optical variability: 1 point-like from \citet{t89} and 20 with diffuse images from BTK.   
For these objects we have computed the  upper limits on the X-ray flux.
Consistency of their X-ray and optical properties with the AGN character is discussed in the next section.

In summary, 21/30=70\% of the variable candidates are not detected in X-rays; however, since 19/44=43\% of known AGNs in our area are not detected in X-rays, we expect that a significant fraction of the variability selected candidates are genuine AGNs, but with low $X/O$.

\section{Results and Discussion}

\begin{figure}
   \centering
\resizebox{\hsize}{!}{\includegraphics{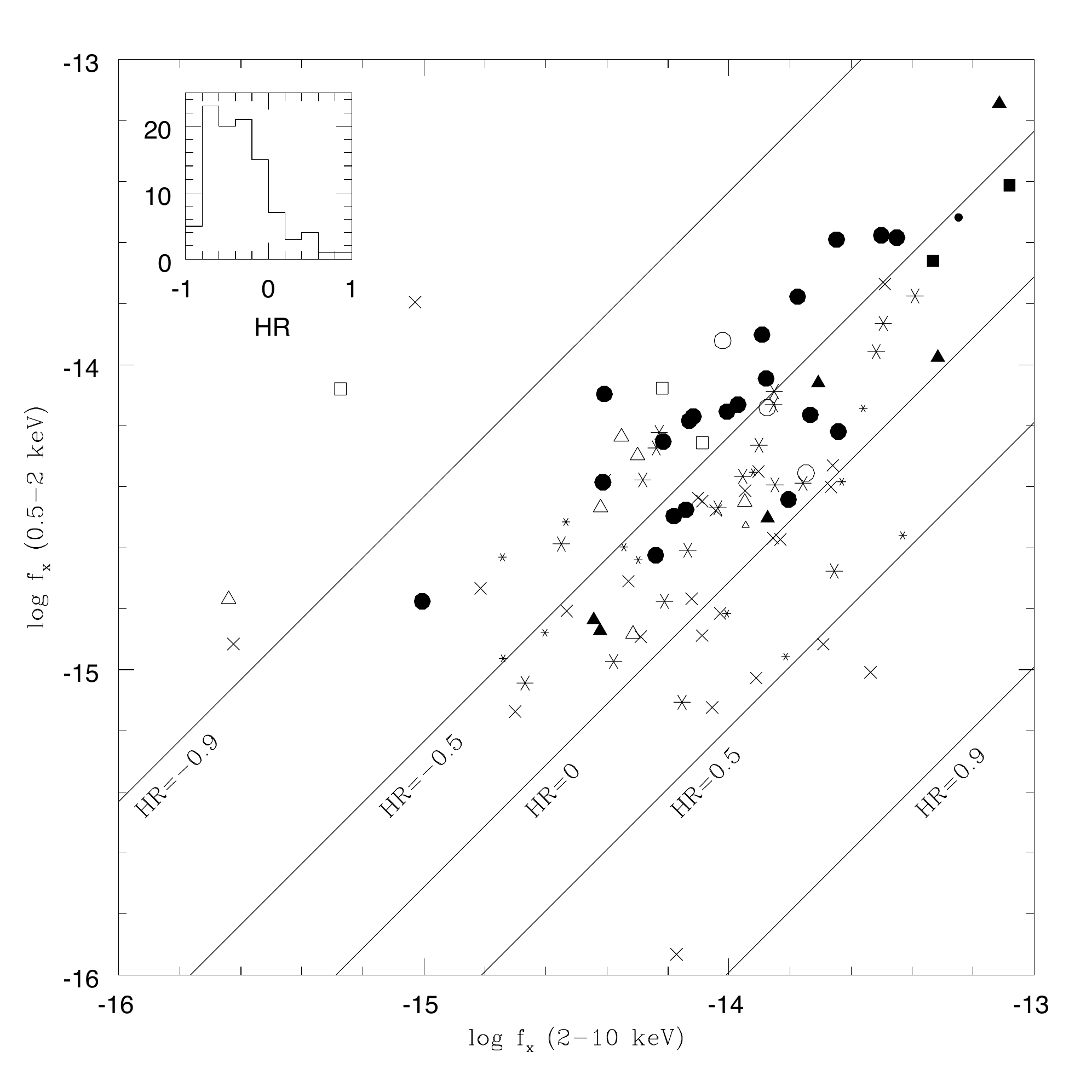}}

      \caption{$f_x(2-10)$ keV (hard band H) versus $f_x(0.5-2)$ keV (soft band S). Solid symbols: objects of known redshift;
open symbols: objects without measured redshift; crosses: optically unidentified sources; 
asterisks: optically identified but unclassified sources; triangles: galaxies; 
circles: point-like AGNs; squares: variability selected AGN candidates with extended images (BTK);
smaller symbols: sources with marginal (M) or ambiguous (A) identifications (see Table \ref{Tab2}).
Continuous lines represent loci of constant hardness ratio $HR$. The inset shows the distribution of hardness ratios.
}
         \label{Fig2}
\end{figure}

Figure \ref{Fig2} shows $f_S$  versus $f_H$. 
Hardness ratios $HR$ have been computed from the count rates $c_S$ and $c_H$ in the soft and hard band:
$HR = (c_H-c_S)/(c_H+c_S)$. All confirmed QSOs/AGNs have $HR$ in the range -0.1, -0.8.
About 10\% of the objects show positive $HR$ values, suggesting absorption in the S band, as expected for Type-2 objects. Most of these hard spectrum objects are optically unidentified.

%%%%%%%%%%%%%% Figure 2%%%%%%%%%%%%%%%%
%%%%%%%%%%%%%%%%%%%%%%%%%%%%%%%%%%%%%%%

\begin{figure}
   \centering
\resizebox{\hsize}{!}{\includegraphics{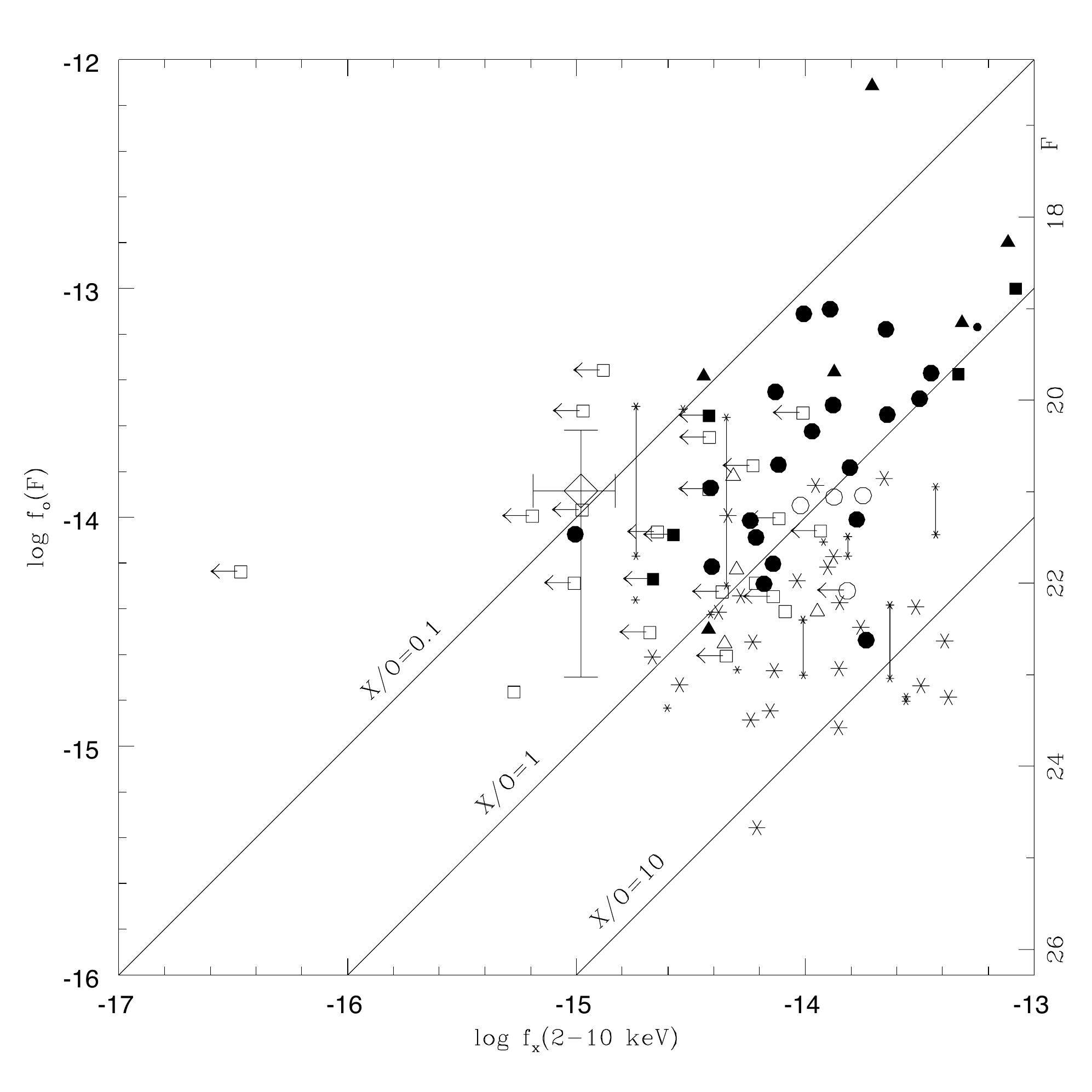}}

      \caption{
Optical F band flux $f_{_F}$ (also shown as apparent magnitudes $F$) versus X-ray (2-10 keV) flux $f_x$(2-10 keV).
Solid symbols: objects of known redshift;
open symbols: objects without measured redshift; crosses: optically unidentified sources; 
asterisks: optically identified but unclassified sources; triangles: galaxies; 
circles: point-like AGNs; squares: variability selected AGN candidates with extended images (BTK);
smaller symbols: sources with marginal (M) or ambiguous (A) identifications (see Table \ref{Tab2}).
Arrows: 1-$\sigma$ upper limits.
Connected symbols represent ambiguous identifications. Diamond with error bars represents the average fluxes
of the X-ray undetected BTK objects, obtained by X-ray image stacking. 
}
         \label{Fig3}
   \end{figure}

Figure \ref{Fig3} shows the flux in the optical F band versus the 2-10
keV flux.  The straight lines represent constant values of the X-ray
(2-10 keV) to optical (F band) flux ratio ($X/O$).  The range of $X/O$
ratio is wide. Considering only the confirmed AGNs with measured X-ray
fluxes, the average $\log(X/O)$ is -0.24 with a standard deviation of
0.4, and corresponds to $X/O=0.58$.  The average $X/O$ found by the
HELLAS2XMM survey is 1.2 with a standard deviation of 0.3
\citep{fior03}.  The difference is mainly due to the different selection technique, since
all the confirmed AGN in SA 57 were selected in the optical band. In
fact the average $X/O$ of a sample of 35 optically selected PG QSOs was
found to be 0.3 with a standard deviation of 0.3 on the basis of ASCA
and BeppoSAX data \citep{fior03,george00,mine00}, consistent with our
findings within the statistical uncertainty.

%%%%%%%%%%%%%% Figure 3%%%%%%%%%%%%%%%%
%%%%%%%%%%%%%%%%%%%%%%%%%%%%%%%%%%%%%%%

In Figure \ref{Fig3}, 9 objects selected on the sole basis of
variability and with measured X-ray flux appear, indicated in the
notes to Table \ref{Tab3}.  They show an $X/O$ consistent with the rest
of the QSO sample.  Three of them (empty circles) are unconfirmed
point-like objects.  We assume the X-ray emission, together with the
$X/O$ typical of other AGNs/QSOs, is a confirmation of their AGN nature.
A fourth variability-selected point-like object (NSER 15248) has also
a spectroscopic confirmation. The other 5 variability-selected objects
with a measured X-ray flux are BTK objects. The optically brightest two
possess optical spectra (filled squares) and are the brightest in
X-rays too. For the fainter three (open squares) again we assume that
X-ray identification and $X/O$ value confirm their AGN nature.

Optically undetected X-ray sources, indicated as crosses in Figure
\ref{Fig2}, are not reported in Figure \ref{Fig3}, where they would
appear as upper limits, $ f_o(F) \sim 2 \times 10^{-15}$ erg cm$^{-2}$
s$^{-1}$, and are discussed below. 

For another 21 X-ray undetected objects selected by optical
variability, one point-like and 20 with extended images (BTK), upper
limits on the 2-10 keV flux are indicated by arrows. While 3-$\sigma$
upper limits would be more secure, they would not be particularly
significant, since most of the objects would result consistent with
typical $X/O$ of AGNs. Therefore we reported in the figure 1-$\sigma$
upper limits, providing smaller $X/O$ ratios. Still, most of these
objects show $0.1<X/O<3$, indicating their consistency with typical
AGNs. A few objects, instead, show upper limits on $X/O$ smaller than
0.1, more consistent with X-ray emitting normal galaxies or star-burst
galaxies.  

As noted in Section 3.2, a large fraction (70\%) of the objects selected by their optical variability is not detected in X-rays. A similar fraction was found by \citet{sara06} in their {\it Chandra/XMM-Newton} and spectroscopic study of the Groth Westphal Survey Strip.
To further investigate possible X-ray emission from our 21 undetected objects, we extracted X-ray
images, in S and H bands, centred on their optical position. We then
produced the corresponding stacked images, thus increasing by a factor
$\sim \sqrt{20}$ the S/N ratio. In both bands a very faint ``object''
is visible in the centre of the stacked image. To verify that the
result is not dominated by one or a few of the individual images, we
have carefully inspected each of them. Only in one case (NSER 10693)
there is a hint of possible photon excess respect to the local
background, contributing 30\% of the flux in the S band and only 1\%
in the H band. After the exclusion of this object the resulting fluxes
in the S and H bands are $1.9\pm0.6\times 10^{-16}$ and $1.0\pm 0.4 \times
10^{-15}$ erg cm$^2$ s$^{-1}$, respectively, with a probability $1.5
\times 10^{-4}$and $4 \times 10^{-4}$, respectively, of being due to
Poisson fluctuations.  The average optical flux in the optical (F)
band is $1.3\pm1.1 \times 10^{-14}$ erg cm$^2$ s$^{-1}$. The
uncertainties in the X-ray fluxes are computed from the ``object'' and
background photon counts in the stacked images; the uncertainty in the
average optical flux is the standard deviation of the 20 measured
fluxes. The corresponding point, representing the average X-ray
undetected BTK object, lies about a factor $\ga$ 2 below the limit of
the present X-ray survey in both the S and H band and has $X/O \sim 8
\times 10^{-2}$. Its hardness ratio is about 0, i.e., relatively high
but consistent with the distribution of AGN hardness ratios.  Thus
this average object could consist of a possibly partially absorbed
AGN, hosted by a galaxy which contributes to the optical flux.

Let us consider the sample which includes all objects in the field with known redshift which are either
i) X-ray detected, i.e., 35 objects with redshift in Table \ref{Tab3}; or ii) confirmed AGNs, 
which were not detected in X-rays, i.e., the 19 objects of Table \ref{Tab4},
plus 3 spectroscopically confirmed BTK objects not detected in X-rays.
For all of them we can compute the optical luminosity and either the X-ray luminosity
or a 3-$\sigma$ upper limit, in the 2-10 keV band.

This  sample  contains 57 objects. Of them, 44 are QSOs,  25 of which  are detected in the X-ray band.
Concerning the 5 extended variable sources (BTK), two are detected in X-rays, a third,
(NSER 4326) falls on  a gap of the EPIC camera, another object (NSER 8553) shows both emission and absorption features and has an 
uncertain AGN characterisation and, lastly, for NSER 16338 the AGN character has been confirmed and the redshift measured in a 
new spectroscopic survey of  SA 57 which is being conducted at  Telescopio Nazionale Galileo (TNG) and  William Herschel Telescope
at La Palma \citep{trev07}. The sample contains also one radio galaxy and
7 objects classified as galaxies in \citet{munn97}.

\begin{figure}
   \centering
\resizebox{\hsize}{!}{\includegraphics{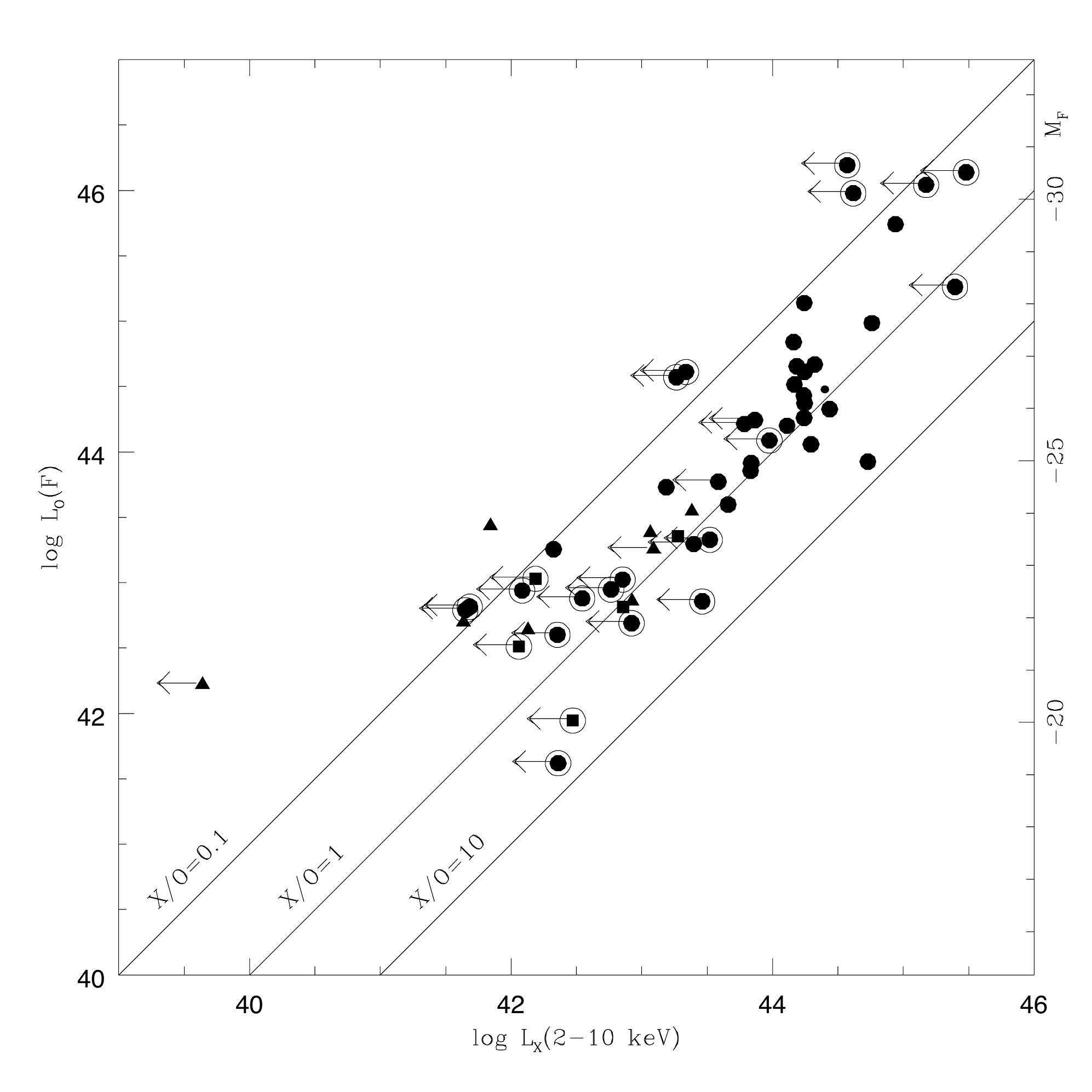}}
      \caption{
Optical F band luminosity $L_o$ versus the X-ray  (2-10 keV) band luminosity $L_x(2-10$ keV$)$.
Luminosities are also shown as absolute magnitudes $M_F$; note that for the objects in figure $\langle B_J-F\rangle \simeq 0.7$. Solid symbols: objects of know redshift; triangles: galaxies; 
circles: point-like AGNs; squares: variability selected AGN candidates with extended images (BTK);
smaller symbols: sources with marginal (M) or ambiguous (A) identifications (see Table \ref{Tab2}).
Arrows represent 3-$\sigma$ (instead of 1-$\sigma$) upper limits: these occur for both X-ray undetected objects from NED and from BTK (which are marked in this figure with big circles) and for objects detected in X-ray bands other than H (without big circles). The continuous lines represent the indicated constant values of the $X/O$ ratio.} 
         \label{Fig4}
\end{figure}

All the above objects are reported in Figure \ref{Fig4}, which shows
the F band luminosity $L_o(F)$ versus the 2-10 keV luminosity
$L_x$(2-10 keV).  In terms of absolute X-ray and optical luminosities,
X-ray detected BTK objects populate the faint ($L_x(2-10 $ keV$) <3
\cdot 10^{43}$ erg s$^{-1}$) side of the diagram as
expected. Objects not detected in the H band are shown in Figure \ref{Fig4} as
3-$\sigma$ upper limits, i.e., as robust but high $X/O$ upper limits.
Some of these objects are detected in at least one of the other X-ray bands, thus they appear in Table \ref{Tab2}. The other upper limits, marked with big circles, correspond to the 22 above mentioned X-ray undetected objects from NED and BTK. The X-ray undetected BTK objects lie close to $L_x(2-10 $ keV$)=
10^{42}$ erg s$^{-1}$, above which the X-ray emission is usually
attributed to an active nucleus rather than star-burst activity.  At
least some of them likely have $X/O < 0.1$.  Thus, selecting
``variable galaxies'', i.e., variable objects with diffuse images, we
are indeed selecting intrinsically faint AGN, whose host galaxy is not
swamped by the nuclear luminosity. The low $X/O$ ratio may be due to the
contribution of the host galaxy to the optical luminosity.

Of the 7 X-ray detected galaxies with redshift known from
\citet{munn97}, one (SA57X 69) is a relatively nearby ($z=0.0213$)
spiral galaxy with $L_x(2-10 $ keV$) \la 4 \times 10^{39}$ erg
s$^{-1}$. Another object is the cD galaxy at the centre of the galaxy
cluster II Zw 1305.4+2941.  Of the remaining 5 objects, 3 have
$L_x(2-10 $ keV$)$ and $X/O$ consistent with those of (faint) AGNs,
while the other two have $X/O <0.1$ and $L_x(2-10 $ keV$) \la 10^{42}$
erg s$^{-1}$, consistent with the luminous tail of the ``normal''
galaxy X-ray luminosity function \citep{georga05}.  

Concerning the X-ray undetected point-like objects, shown in Figure
\ref{Fig4} as 3-$\sigma$ upper limits: six (i.e., ~15\%) exhibit $X/O < 0.1$.
This can be compared with ~1\%
found in X-ray selected samples \citep[see][]{laor97,fior03}. We
stress that at least two of them have $L_o(F) \ga 10^{46}$ erg
s$^{-1}$, so that the low $X/O$ cannot be due to the contribution of the
host galaxy to the optical light, as in the case of fuzzy objects, but
must be intrinsic to the nuclear component.

%#######
%##Figure 4
%#######

%#######
%##Figure 5
%#######
\begin{figure}
   \centering
\resizebox{\hsize}{!}{\includegraphics{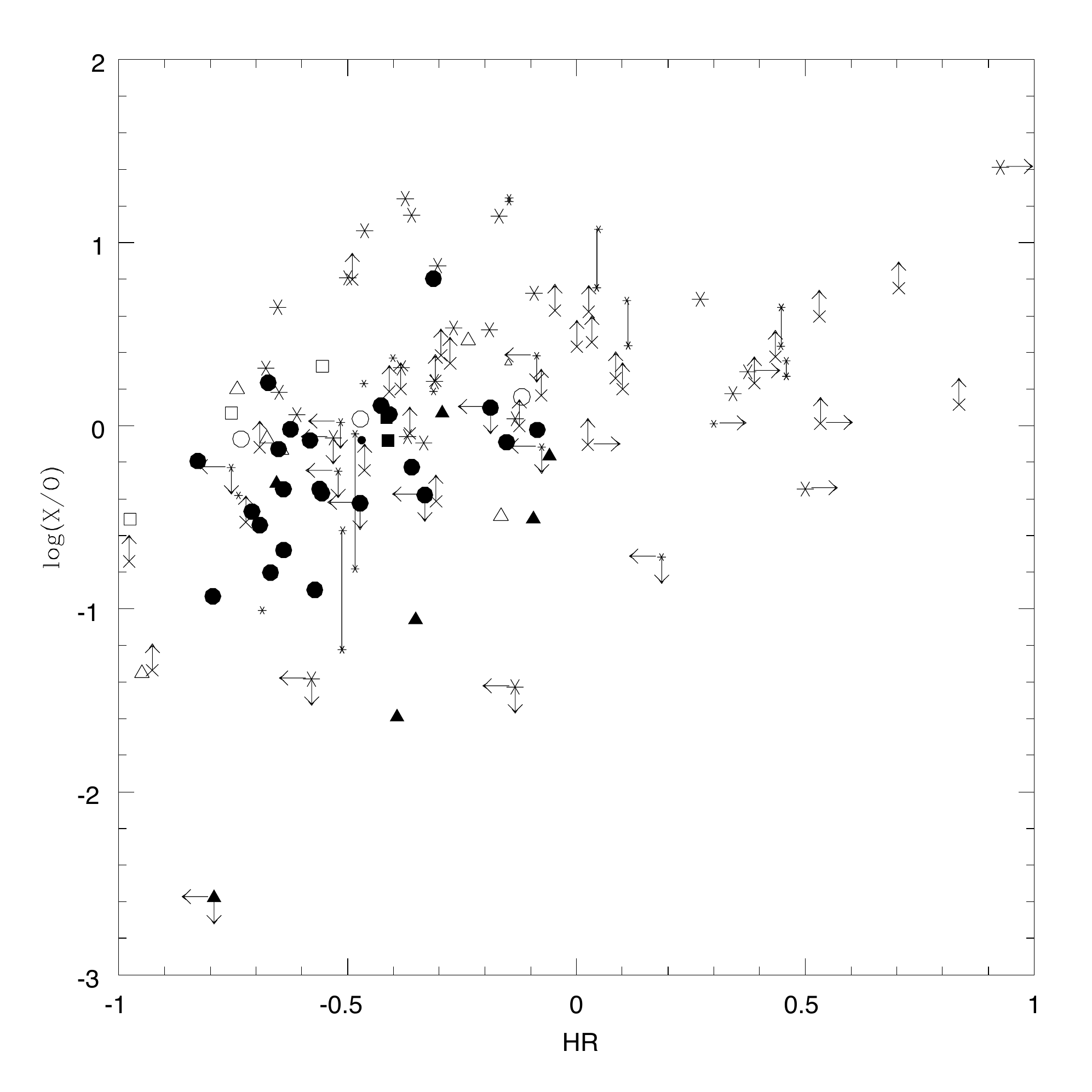}}
      \caption{$X/O$ versus hardness ratio $HR$. 
 Solid symbols: objects of know redshift; open symbols: objects without measured redshift; 
crosses: optically unidentified sources; 
asterisks: optically identified but unclassified sources; triangles: galaxies; 
circles: point-like AGNs; squares: variability selected AGN candidates with extended images (BTK);
smaller symbols: sources with marginal (M) or ambiguous (A) identifications (see Table \ref{Tab2}), 
connected in the case of ambiguous identifications. Horizontal and downward vertical arrows: 3-$\sigma$ upper limits derived from X-ray bands; upward vertical arrows: optical limit $F>22$.}
         \label{Fig5}
\end{figure}

In Figure \ref{Fig5} we report the $X/O$ ratio versus the hardness
ratio $HR$ for all of the objects detected in at least two of the S,
H, T bands. Horizontal and downward vertical arrows are also reported to indicate
X-ray limits when the relevant detections are missing.
The X-ray detected objects not seen in the optical F
band have $X/O$ ratios reported as lower limits assuming $F=22$ as the
limiting magnitude (where the optical sample is $\sim$ 50\% complete).
Some of these objects have higher $X/O$ in comparison to the rest of the
sample, consistent with being obscured in the optical and soft X-ray bands, but not in
hard X-rays.

While optically undetected objects (crosses) are spread over the entire $HR$ range, their fraction is higher among the hard (e.g. $HR>0$) objects. This is consistent with the analysis of \citet{szok04}, which is deeper than ours both in X-ray and optical bands, and shows that hard X-ray objects have on average fainter optical magnitudes.

Thus, objects lying in the upper right of the figure, e.g., $X/O \ga 3$ and
$HR \ga 0$, are good type-2 AGN candidates, where the nuclear component
is obscured and the colour is dominated by the light from the host
galaxy.  Such objects are not selected as AGN candidates from optical
observations (i.e., on the basis of non-stellar colours or
variability), while they are detected in the X-ray band.  The other
optically identified objects (asterisks) require optical spectroscopy
to discriminate among starburst galaxies, AGNs or normal X-ray
emitting galaxies.

\begin{figure}
   \centering
\resizebox{\hsize}{!}{\includegraphics{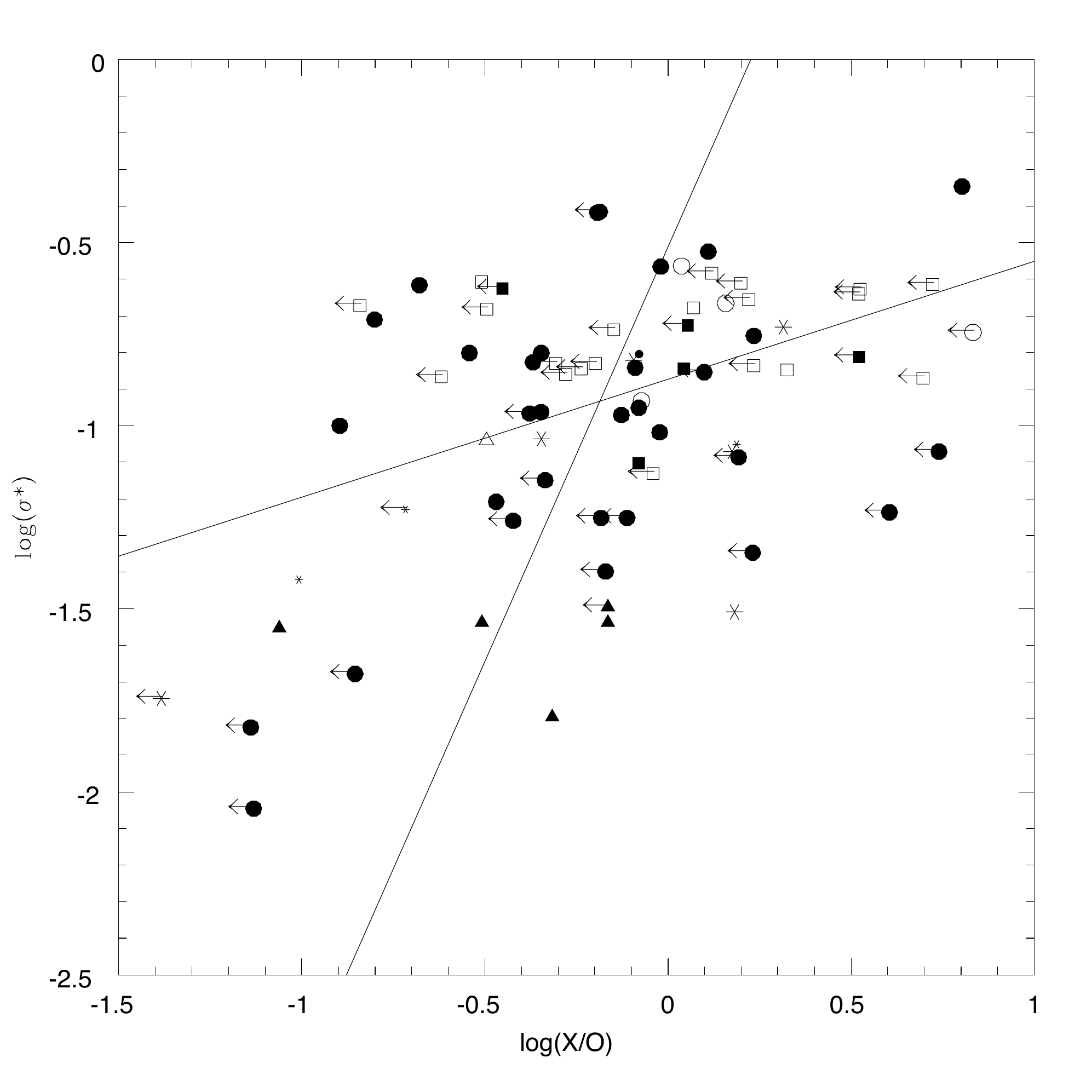}}
      \caption{
Logarithm of the optical variability, as measured by the the r.m.s. magnitude changes $\sigma^\ast$, versus the logarithm of the $X/O$ ratio.
Symbols as in the previous figures. Arrows: 3-$\sigma$ upper limits.The regression lines shown are computed excluding the upper limits.}
         \label{Fig6}
\end{figure}

In Figure \ref{Fig6} we report the variability amplitude  $\sigma^\ast$ measured by the r.m.s. magnitude changes \citep{t89,btk98} versus the $X/O$ ratio. Point-like objects were classified as variable for $\sigma\ast>0.1$ mag, while the variability threshold for extended objects was in the range 0.06-0.2 mag, depending on the object magnitude \citep{t89,btk98}. Data in Fig. 6 are reported both for variable and non variable objects.

For the objects with measured $X/O$ a Pearson correlation coefficient 
$r = 0.38$ is found, with a probability of the null hypothesis $P(>r) = 0.02$. 

%#######
%##Figure 6
%#######

Since the fraction of objects with only upper limits on $X/O$ is about
50\%, we also computed both the Kendall $\tau$ and the Spearman $\rho$
rank correlation coefficients as generalised for application to
censored data \citep[see e.g.][]{isob86,akri96}) adopting 3-$\sigma$
upper limits on $X/O$.  The results is: $\tau=0.26$ with
$P(>\tau)=0.011$ and $\rho=0.31$ with $P(>\rho)=0.007$. To check to
what extent the above results rely on the most deviant point on the
top right of Figure \ref{Fig6} (SA57X 33, a quasar at $z=2.124$), we
re-evaluated the rank correlations after removing this object from the
list. The results ($\tau'=0.22$ with $P(>\tau')=0.03$ and $\rho'=0.27$
$P(>\rho')=0.017$) do not change substantially and remain moderately
significant (a 2.4-$\sigma$ result if the probability distribution were Gaussian).  We note that, on average, optical AGN variability
decreases with luminosity as $\sigma^\ast \propto L^{-0.25}$
\citep{vand04}, while the X-ray to optical ratio decreases with
luminosity as $X/O \propto L^{-\chi}$ with $\chi \approx 0.25-0.45$
\citep{stra05,vign06,miya06}. It is still unclear whether the
$X/O$-luminosity anti-correlation reflects some physical property of the
AGNs or is simply due to a selection effect.  In both cases this
correlation is qualitatively consistent with our results showing an
increase of variability with the X-ray to optical ratio.  This
suggests to further investigate the $X/O$-$\sigma^\ast$ correlation
and its possible physical origin.

\section{Summary}
\begin{itemize}
\item{We have surveyed with XMM-Newton a $\sim$0.2 deg$^2$ area in the field of SA 57 down to flux limits of $10^{-15}$,  $5 \times 10^{-16}$, $2 \times 10^{-15}$, $10^{-14}$  erg cm$^{-2}$  s$^{-1}$ in the (0.5-10 keV) (T), (0.5-2 keV) (S), (2-10 keV) (H), and (5-10 keV) (HH) band respectively.} 
\item{Adopting a threshold probability  of $2 \times 10^{-5}$ of detecting Poisson fluctuations of the Background,
we found  119, 117, 58, 6 sources in the T, S, H, and HH bands respectively.}
\item{The distribution of X-ray to optical flux ratios shows an excess of low $X/O$ among optically selected AGNs
in agreement with previous studies of optically selected PG QSOs \citep{fior03}.}
\item{We detect X-ray emission from 9 objects selected on the sole basis of variability. We assume this is
a confirmation of their AGN character. Three of them (1 point-like and 2 extended)  also possess  
optical spectra consistent with their AGN nature.}
\item{Variability selected, but X-ray undetected objects show 1-$\sigma$ upper limits on $X/O$ consistent with
the AGN population. The stacked image of these objects indicates  average X-ray fluxes of
$1.9\pm0.6 \times10^{-15}$ and  $1.0\pm0.4 \times 10^{-17}$ erg cm$^2$ s$^{-1}$  in the S and H bands respectively,
significant at more than 3-$\sigma$ level, corresponding to a low $X/O \sim 8\times 10^{-2}$ and $HR\sim 0$.}
\item{Of the known QSOs in the field only 25/44 are detected in X-rays.}
\item{On the basis of 3-$\sigma$ upper limits, 15\% of QSOs have $X/O < 0.1$, compared to only 1\% of X-ray selected samples.}
\item{Low $X/O$ in objects with $L_o(F) \ga 10^{46}$ erg s$^{-1}$
must be intrinsic of the nuclear component rather than due the host galaxy light.}
\item{Most X-ray selected objects, not previously identified as AGNs, are probably type-2 objects dominated by the stellar components.}
\item{The optical variability appears marginally  correlated with the $X/O$ ratio. This could be explained by the decrease with luminosity of both $X/O$ and variability.}
\item{From this survey we will derive in a forthcoming paper an estimate of the space density of LLAGNs after further progress of the ongoing spectroscopic survey of SA57 \citep{trev07}.}
\end{itemize}

\begin{acknowledgements}
We thank Valentina Zitelli for useful discussions. We thank the referee, Vicki Sarajedini, for remarks and suggestions. We acknowledge partial support of Agenzia Spaziale Italiana and Istituto Nazionale di Astrofisca by the grant ASI/INAF n. I/023/05/0.
This research has made use of the NASA/IPAC Extragalactic Database (NED) which is operated by the Jet Propulsion Laboratory, 
California Institute of Technology, under contract with the National Aeronautics and Space Administration.
\end{acknowledgements}

\begin{longtable}{ccccccc}
\caption{\label{Tab2} The source catalog}\\
\hline\hline
SA57X &  RA (2000) &  DEC (2000) & \multicolumn{3}{c}{flux [$10^{-14}$ erg/cm$^2$/s]} &  IdRank$^{\rm a}$ \\
 &   &   & 0.5-10 keV  &  0.5-2 keV  & 2-10 keV &  \\
\hline
\endfirsthead
\caption{continued}\\
\hline\hline
SA57X &  RA (2000) &  DEC (2000) & \multicolumn{3}{c}{flux [$10^{-14}$ erg/cm$^2$/s]} &  IdRank$^{\rm a}$ \\
 &   &   & 0.5-10 keV  &  0.5-2 keV  & 2-10 keV &  \\
\hline
\endhead
\hline
\endfoot
  1 & 13 08 11.15 & 29 10 50.4 &    0.50    &    0.50    & $<$0.44     &  U \\
  2 & 13 08 38.94 & 29 11 22.4 &    0.13$^{\rm b}$  &    0.13    & $<$0.41     &  U \\
  3 & 13 07 55.57 & 29 11 26.2 &    0.34    &    0.19    &    0.15$^{\rm b}$   &  U \\
  4 & 13 08 56.66 & 29 11 58.3 &    1.64    &    0.11    &    1.53$^{\rm b}$   &  A \\
  5 & 13 08 44.75 & 29 12 14.4 &    2.16    &    0.12    &    2.04$^{\rm b}$   &  U \\
  6 & 13 08 33.60 & 29 12 20.4 &    3.01    &    0.10$^{\rm b}$  &    2.91     &  U \\
  7 & 13 08 51.78 & 29 12 34.3 &    2.66    &    0.47    &    2.19     &  U \\
  8 & 13 08 49.33 & 29 12 42.3 &    1.40    &    0.66    &    0.74     &  I \\
  9 & 13 07 54.29 & 29 12 50.2 &    5.90    &    1.06    &    4.84     &  I \\
 10 & 13 08 50.86 & 29 13 06.3 &    0.64    &    0.13    &    0.51$^{\rm b}$   &  U \\
 11 & 13 08 52.39 & 29 13 26.3 &    0.94    &    0.42    &    0.52$^{\rm b}$   &  I \\
 12 & 13 08 35.43 & 29 13 38.4 &    0.25$^{\rm b}$  &    0.25    & $<$0.39     &  I \\
 13 & 13 08 58.35 & 29 14 02.2 &    4.14    &    1.10    &    3.04     &  I \\
 14 & 13 08 29.17 & 29 14 02.5 &    1.24    & $<$0.07    & $<$0.73     &  M \\
 15 & 13 07 48.67 & 29 14 18.1 &    1.04    & $<$0.09    & $<$0.64     &  I \\
 16 & 13 07 42.11 & 29 14 10.0 &    0.19$^{\rm b}$  &    0.19    & $<$0.42     &  U \\
 17 & 13 07 41.80 & 29 14 22.0 &    4.57    &    1.37    &    3.20$^{\rm b}$   &  I \\
 18 & 13 08 59.42 & 29 14 30.2 &    5.08    &    1.84    &    3.24     &  U \\
 19 & 13 07 37.36 & 29 14 51.9 &    2.14    &    0.74    &    1.40$^{\rm b}$   &  I \\
 20 & 13 09 00.34 & 29 14 54.2 &    0.81    &    0.24    &    0.58$^{\rm b}$   &  I \\
 21 & 13 08 12.05 & 29 14 58.4 &    3.35    &    1.67    &    1.68     &  I \\
 22 & 13 09 20.51 & 29 14 57.8 &    1.33    & $<$0.07    & $<$0.44     &  I \\
 23 & 13 08 27.03 & 29 15 22.5 &    2.88    &    0.60    &    2.28     &  I \\
 24 & 13 08 26.72 & 29 15 30.5 &    0.41$^{\rm b}$  & $<$0.07    &    0.41     &  U \\
 25 & 13 08 18.47 & 29 15 30.4 &    2.23    &    0.44    &    1.79     &  I \\
 26 & 13 08 45.37 & 29 15 30.4 &    0.15    &    0.12    &    0.02$^{\rm b}$   &  U \\
 27 & 13 09 11.96 & 29 15 54.0 &    5.81    &    2.66    &    3.16     &  I \\
 28 & 13 08 17.86 & 29 16 06.4 &    1.70    &    0.45    &    1.25     &  U \\
 29 & 13 08 15.72 & 29 16 10.4 &    1.93    &    0.36    &    1.57$^{\rm b}$   &  I \\
 30 & 13 08 27.95 & 29 16 18.5 &    0.22$^{\rm b}$  &    0.22    & $<$0.36     &  M \\
 31 & 13 07 38.11 & 29 16 25.9 &    6.15    &    2.61    &    3.55     &  I \\
 32 & 13 08 34.06 & 29 16 46.4 &    0.81    &    0.42    &    0.39     &  U \\
 33 & 13 09 04.64 & 29 17 30.1 &    2.53    &    0.68    &    1.85     &  I \\
 34 & 13 08 47.51 & 29 17 46.4 &    1.14    &    0.15    &    0.98$^{\rm b}$   &  A \\
 35 & 13 07 49.65 & 29 17 51.1 &    1.37    &    0.56    &    0.82$^{\rm b}$   &  I \\
 36 & 13 09 14.12 & 29 17 58.0 &    0.98    & $<$0.08    & $<$0.57     &  U \\
 37 & 13 08 22.14 & 29 18 06.4 &    0.46$^{\rm b}$  & $<$0.03    &    0.46     &  I \\
 38 & 13 08 13.49 & 29 18 09.4 &    0.54    &    0.26    &    0.28$^{\rm b}$   &  I \\
 39 & 13 08 35.90 & 29 18 22.4 &    0.95    &    0.13    &    0.82$^{\rm b}$   &  U \\
 40 & 13 07 57.97 & 29 18 42.2 &    0.52    &    0.11    &    0.42$^{\rm b}$   &  I \\
 41 & 13 08 39.87 & 29 18 50.4 &    0.32    &    0.32    & $<$0.82     &  I \\
 42 & 13 08 32.23 & 29 19 10.4 &    1.81    &    0.74    &    1.07     &  I \\
 43 & 13 08 44.77 & 29 19 10.4 &    0.22    & $<$0.07    & $<$0.54     &  U \\
 44 & 13 08 42.85 & 29 19 15.4 &    0.41    &    0.23    &    0.18$^{\rm b}$   &  M \\
 45 & 13 08 28.86 & 29 19 14.4 &    0.09$^{\rm b}$  &    0.09    & $<$0.39     &  M \\
 46 & 13 08 02.87 & 29 19 22.3 &    0.39$^{\rm b}$  & $<$0.04    &    0.39     &  M \\
 47 & 13 08 16.02 & 29 19 18.4 &    1.69    &    0.70    &    0.98     &  I \\
 48 & 13 07 42.68 & 29 19 34.0 &    0.98    &    0.32    &    0.66$^{\rm b}$   &  I \\
 49 & 13 08 32.23 & 29 19 54.4 &    1.80    &    0.55    &    1.25     &  I \\
 50 & 13 09 10.39 & 29 20 03.0 &    0.64    &    0.64    & $<$0.81     &  U \\
 51 & 13 08 01.64 & 29 20 10.3 &    0.27    &    0.17    &    0.10$^{\rm b}$   &  I \\
 52 & 13 08 22.13 & 29 20 10.4 &    5.75    &    1.68    &    4.07     &  I \\
 53 & 13 07 41.14 & 29 20 17.9 &    2.15    &    0.41    &    1.75     &  I \\
 54 & 13 07 45.12 & 29 20 22.0 &    0.73    &    0.23    &    0.50$^{\rm b}$   &  M \\
 55 & 13 08 12.80 & 29 20 38.4 &    0.96    &    0.08$^{\rm b}$  &    0.88     &  U \\
 56 & 13 08 47.53 & 29 20 46.4 &    1.64    &    0.44    &    1.20     &  M \\
 57 & 13 09 00.98 & 29 20 46.2 &    0.50    & $<$0.04    & $<$0.47     &  M \\
 58 & 13 08 45.38 & 29 20 58.4 &    0.15    & $<$0.06    & $<$0.63     &  U \\
 59 & 13 07 15.74 & 29 21 01.2 &    0.98$^{\rm b}$  &    0.98    & $<$0.40     &  U \\
 60 & 13 07 58.88 & 29 21 06.3 &    2.42    &    0.21    &    2.21     &  I \\
 61 & 13 08 55.48 & 29 21 06.3 &    0.23    &    0.23    & $<$0.32     &  I \\
 62 & 13 08 40.18 & 29 21 10.4 &    0.15    &    0.15    & $<$0.64     &  M \\
 63 & 13 08 30.39 & 29 21 26.5 &    1.16    &    0.37    &    0.79     &  U \\
 64 & 13 09 15.44 & 29 21 26.9 &    0.68    &    0.01$^{\rm b}$  &    0.67     &  U \\
 65 & 13 08 56.09 & 29 21 26.3 &    0.08$^{\rm b}$  &    0.08    & $<$0.25     &  U \\
 66 & 13 08 56.40 & 29 21 46.3 &    0.06$^{\rm b}$  &    0.06    & $<$0.43     &  A \\
 67 & 13 07 47.33 & 29 21 39.1 &    1.16    & $<$0.09    & $<$0.71     &  M \\
 68 & 13 08 57.62 & 29 21 54.2 &    0.24$^{\rm b}$  &    0.24    & $<$0.39     &  A \\
 69 & 13 09 15.91 & 29 22 02.9 &    0.82    &    0.82    & $<$0.49     &  I \\
 70 & 13 08 26.41 & 29 22 22.5 &    1.32    &    0.09$^{\rm b}$  &    1.23     &  U \\
 71 & 13 08 56.40 & 29 22 34.3 &    0.60    &    0.31    &    0.29$^{\rm b}$   &  M \\
 72 & 13 08 57.63 & 29 22 42.2 &    0.07$^{\rm b}$  &    0.07    & $<$0.40     &  U \\
 73 & 13 08 23.35 & 29 22 34.5 &    1.17    &    0.56    &    0.61     &  I \\
 74 & 13 08 46.31 & 29 22 34.4 &    0.45    &    0.16    &    0.29$^{\rm b}$   &  U \\
 75 & 13 08 49.67 & 29 22 38.3 &    0.66    &    0.20    &    0.47$^{\rm b}$   &  U \\
 76 & 13 08 06.22 & 29 22 38.3 &    1.19    &    0.80    &    0.39     &  I \\
 77 & 13 08 13.26 & 29 22 44.4 &    2.22    &    0.90    &    1.32     &  I \\
 78 & 13 08 57.94 & 29 22 50.2 &    0.08$^{\rm b}$  &    0.08    & $<$0.32     &  U \\
 79 & 13 08 37.13 & 29 22 50.4 &    0.72    &    0.34    &    0.38     &  I \\
 80 & 13 09 08.11 & 29 22 51.1 &    0.98    &    0.25    &    0.73$^{\rm b}$   &  I \\
 81 & 13 09 00.08 & 29 22 58.2 &    0.13    &    0.13    & $<$0.50     &  I \\
 82 & 13 07 33.47 & 29 23 01.8 &    6.85    &    2.19    &    4.66     &  I \\
 83 & 13 08 51.82 & 29 23 10.3 &    0.78    &    0.08$^{\rm b}$  &    0.70     &  I \\
 84 & 13 08 04.99 & 29 23 22.3 &    1.18    &    0.36    &    0.82     &  U \\
 85 & 13 07 44.79 & 29 23 28.0 &    1.09    &    0.15    &    0.94$^{\rm b}$   &  U \\
 86 & 13 08 19.38 & 29 23 34.4 &    0.19    &    0.17    &    0.02$^{\rm b}$   &  I \\
 87 & 13 08 37.13 & 29 23 34.4 &    0.93    &    0.17    &    0.75     &  U \\
 88 & 13 08 16.31 & 29 23 54.4 &    0.51    &    0.15    &    0.36     &  I \\
 89 & 13 07 57.41 & 29 24 09.2 &    2.76    &    0.41    &    2.34     &  A \\
 90 & 13 08 49.99 & 29 24 14.3 &    0.29    &    0.11    &    0.18$^{\rm b}$   &  A \\
 91 & 13 08 27.64 & 29 24 22.5 &    2.83    &    0.87    &    1.96$^{\rm b}$   &  I \\
 92 & 13 08 12.64 & 29 24 34.4 &    1.26    &    0.34    &    0.92     &  I \\
 93 & 13 08 40.49 & 29 24 48.4 &    1.24    &    0.33    &    0.91     &  U \\
 94 & 13 08 47.54 & 29 24 50.4 &    1.11    &    0.53    &    0.58     &  I \\
 95 & 13 08 30.39 & 29 25 02.5 &    0.17    &    0.17    & $<$0.61     &  I \\
 96 & 13 08 11.72 & 29 25 14.4 &    2.54    &    1.25    &    1.28     &  I \\
 97 & 13 08 25.19 & 29 25 22.5 &    0.78    &    0.17    &    0.61     &  I \\
 98 & 13 08 50.29 & 29 25 38.3 &    0.13$^{\rm b}$  &    0.13    & $<$0.50     &  U \\
 99$^{\rm c}$& 13 07 49.37 & 29 25 50.1 &   14.85    &    7.17    &    7.68     &  I \\
100 & 13 07 42.01 & 29 26 22.0 &    1.01    &    0.50    &    0.50$^{\rm b}$   &  I \\
101 & 13 08 09.27 & 29 26 26.4 &    1.48    &    0.35    &    1.13     &  I \\
102 & 13 07 29.15 & 29 26 25.6 &    1.54    &    0.43    &    1.11$^{\rm b}$   &  I \\
103 & 13 08 34.38 & 29 26 26.4 &    1.65    &    0.31    &    1.34     &  I \\
104 & 13 08 15.08 & 29 26 46.4 &    0.38    &    0.13    &    0.25$^{\rm b}$   &  M \\
105 & 13 08 10.26 & 29 26 51.4 &    0.62    &    0.13    &    0.48$^{\rm b}$   &  I \\
106 & 13 08 36.52 & 29 26 54.4 &    1.02    &    0.58    &    0.44$^{\rm b}$   &  I \\
107 & 13 09 23.99 & 29 27 17.7 &    3.98    &    0.28$^{\rm b}$  &    3.71     &  A \\
108 & 13 08 08.65 & 29 27 26.4 &    2.06    &    0.72    &    1.34     &  I \\
109 & 13 07 56.71 & 29 27 30.2 &    0.98    & $<$0.05    & $<$0.63     &  M \\
110 & 13 08 13.24 & 29 27 42.4 &    0.31    &    0.09    &    0.21$^{\rm b}$   &  I \\
111 & 13 09 06.53 & 29 27 46.1 &    0.12$^{\rm b}$  &    0.12    & $<$0.22     &  I \\
112 & 13 08 45.40 & 29 27 50.4 &    0.27    &    0.07    &    0.20$^{\rm b}$   &  U \\
113 & 13 08 53.98 & 29 27 50.3 &    1.43    &    0.30    &    1.13     &  M \\
114 & 13 07 51.80 & 29 27 52.2 &    1.44    &    0.84    &    0.60$^{\rm b}$   &  I \\
115 & 13 08 23.96 & 29 28 18.5 &    1.19    &    0.60    &    0.59     &  I \\
116 & 13 08 44.18 & 29 28 42.4 &    0.80    &    0.41    &    0.39$^{\rm b}$   &  I \\
117 & 13 08 45.10 & 29 28 46.4 &    1.33$^{\rm b}$  & $<$0.12    &    1.33     &  I \\
118 & 13 08 55.21 & 29 28 50.3 &    0.70    &    0.25    &    0.45$^{\rm b}$   &  A \\
119 & 13 08 47.24 & 29 29 06.4 &    1.52    &    0.39    &    1.13     &  U \\
120 & 13 08 03.44 & 29 29 10.3 &   12.17    &    3.87    &    8.30     &  I \\
121 & 13 08 21.51 & 29 29 08.4 &    1.70    &    1.60    &    0.09$^{\rm b}$   &  U \\
122 & 13 08 51.23 & 29 29 46.3 &    2.22    &    0.82    &    1.40     &  I \\
123 & 13 08 20.28 & 29 29 58.4 &    0.45    &    0.45    & $<$0.32     &  A \\
124 & 13 08 12.01 & 29 30 26.4 &    3.48    &    0.72    &    2.76     &  A \\
125 & 13 08 24.27 & 29 30 38.5 &    1.44    &    0.68    &    0.76$^{\rm b}$   &  I \\
126 & 13 08 42.04 & 29 30 42.4 &    1.06    &    0.33    &    0.72$^{\rm b}$   &  I \\
127 & 13 09 17.36 & 29 31 02.9 &    2.15    &    1.20    &    0.95$^{\rm b}$   &  I \\
128 & 13 08 08.64 & 29 31 06.4 &    0.53$^{\rm b}$  & $<$0.03    &    0.53     &  U \\
129 & 13 08 48.48 & 29 31 22.4 &    1.66    &    0.27    &    1.39$^{\rm b}$   &  U \\
130 & 13 08 26.41 & 29 31 46.5 &    2.56    &    0.40    &    2.16     &  U \\
131 & 13 08 31.93 & 29 32 22.4 &    1.82    &    0.40    &    1.41     &  I \\
132 & 13 09 11.16 & 29 32 40.0 &    0.27$^{\rm b}$  &    0.27    & $<$0.18     &  U \\
133$^{\rm d}$ & 13 09 05.96 & 29 33 02.1 & $<$0.25    & $<$0.03    & $<$0.61     &  U \\
134 & 13 08 49.56 & 29 33 02.3 &    4.22$^{\rm b}$  & $<$0.03    &    4.22     &  I \\
135 & 13 07 49.93 & 29 33 06.1 &    8.69    &    3.04    &    5.65$^{\rm b}$   &  M \\
136 & 13 08 06.49 & 29 33 34.3 &    0.89    &    0.83    &    0.05$^{\rm b}$   &  I \\
137 & 13 08 47.26 & 29 33 54.4 &    1.13    & $<$0.07    & $<$0.60     &  A \\
138 & 13 07 49.01 & 29 34 22.1 &    1.74    &    0.27    &    1.47$^{\rm b}$   &  U \\
139 & 13 08 33.16 & 29 34 26.4 &    0.51    &    0.13    &    0.38$^{\rm b}$   &  I \\
140 & 13 08 08.32 & 29 34 50.4 &    4.82    &    2.57    &    2.25$^{\rm b}$   &  I \\
\end{longtable}
\noindent
$^{\rm a}$ Rank of optical identification: U=unidentified, I=identified, M=marginal identification, A=ambiguous identification\\
$^{\rm b}$ Object not detected in this band, with flux computed from the other bands\\
$^{\rm c}$ At the same position  an extended source is detected, corresponding to the galaxy cluster ZwCl 1305.4+2941, 
 not considered in the present analysis\\
$^{\rm d}$ This source is only detected in the 5-10 keV band, with $f_x=3.48\times 10^{-14}$ erg/cm$^2$/s

\begin{longtable}{ccccccccc}
\caption{\label{Tab3} Optical identifications}\\
\hline\hline
SA57X & IdRank$^{\rm a}$ & NSER$^{\rm b}$ & RA(2000) & DEC(2000) & $B_J$ & $F$ & $z$ & class$^{\rm c}$\\ 
\hline
\endfirsthead
\caption{continued}\\
\hline\hline
SA57X & IdRank$^{\rm a}$ & NSER$^{\rm b}$ & RA(2000) & DEC(2000) & $B_J$ & $F$ & $z$ & class$^{\rm c}$\\ 
\hline
\endhead
\hline
\endfoot
  4 &  A  & 3049           &  13 08 56.66  &  29 11 58.9   & 23.14 & 21.70 &       &     \\ 
    &     & 3017           &  13 08 56.68  &  29 11 57.3   & 23.03 & 21.49 &       &     \\ 
  8 &  I  & 3544           &  13 08 49.65  &  29 12 44.8   & 20.17 & 19.91 & 1.808 &  Q  \\ 
  9 &  I  & 3535           &  13 07 54.64  &  29 12 46.3   & 20.41 & 19.16 & 0.403 &  G  \\ 
 11 &  I  & 3927           &  13 08 52.56  &  29 13 26.6   & 22.82 & 22.14 &       &     \\ 
 12 &  I  & 3993           &  13 08 35.39  &  29 13 34.6   & 22.95 & 22.14 &       &     \\ 
 13 &  I  & 4274           &  13 08 58.44  &  29 14 02.6   & 23.82 & 22.26 &       &     \\ 
 14 &  M  & 4290           &  13 08 29.43  &  29 14 04.0   & 24.68 &       &       &     \\ 
 15 &  I  & 4437           &  13 07 48.90  &  29 14 17.8   & 22.13 & 21.35 & 0.706 &  G  \\ 
 17 &  I  & 4455           &  13 07 41.86  &  29 14 19.2   & 23.41 & 23.12 &       &     \\ 
 19 &  I  & 4793           &  13 07 37.48  &  29 14 52.4   & 23.91 & 23.58 &       &     \\ 
 20 &  I  & 4811           &  13 09 00.38  &  29 14 53.8   & 22.80 & 21.32 & 3.543 &  Q  \\ 
 21 &  I  & 4882           &  13 08 12.20  &  29 14 59.1   & 21.89 & 21.30 & 1.468 &  Q  \\ 
 22 &  I  & 4855           &  13 09 20.52  &  29 14 58.6   & 22.35 & 21.70 & 1.305 &  Q  \\ 
 23 &  I  & 5141           &  13 08 27.22  &  29 15 25.2   & 20.60 & 20.16 & 1.094 &  Q  \\ 
 25 &  I  & 5185           &  13 08 18.43  &  29 15 30.0   & 21.66 & 21.04 &       & [Q]$^e$ \\ 
 27 &  I  & 5422           &  13 09 11.88  &  29 15 52.7   & 20.62 & 19.98 & 1.083 &  Q  \\ 
 29 &  I  & 5643           &  13 08 15.75  &  29 16 12.3   & 21.17 & 20.74 & 0.983 &  Q  \\ 
 30 &  M  & 5711           &  13 08 28.37  &  29 16 19.0   & 23.78 & 22.45 &       &     \\ 
 31 &  I  & 5767           &  13 07 38.11  &  29 16 26.5   & 20.60 & 19.71 & 0.708 &  Q  \\ 
 33 &  I  & 6442           &  13 09 04.77  &  29 17 29.6   & 22.28 & 22.62 & 2.124 &  Q  \\ 
 34 &  A  & 6579           &  13 08 47.67  &  29 17 44.6   & 23.35 & 23.01 &       &     \\ 
    &     & 6593           &  13 08 47.27  &  29 17 46.2   & 24.51 & 22.40 &       &     \\ 
 35 &  I  & 6669           &  13 07 49.58  &  29 17 54.4   & 22.96 & 22.31 &       & BTK$^e$ \\ 
 37 &  I  & 6825           &  13 08 22.19  &  29 18 07.8   & 22.92 & 21.26 &       &     \\ 
 38 &  I  & 6884           &  13 08 13.52  &  29 18 12.5   & 20.83 & 19.01 &       &     \\
 40 &  I  & 7251           &  13 07 58.01  &  29 18 42.8   & 23.20 & 22.32 &       &     \\ 
 41 &  I  & 7326           &  13 08 39.78  &  29 18 50.2   & 21.03 & 20.55 & 1.315 &  Q  \\ 
 42 &  I  & 7567           &  13 08 32.10  &  29 19 12.0   & 20.51 & 20.34 & 1.812 &  Q  \\ 
 44 &  M  & 7585           &  13 08 42.20  &  29 19 14.4   & 25.40 & 22.18 &       &     \\ 
 45 &  M  & 7727           &  13 08 28.63  &  29 19 26.2   & 22.78 & 22.36 &       &     \\ 
 46 &  M  & 7598           &  13 08 02.56  &  29 19 14.9   & 23.68 & 22.34 &       &     \\ 
 47 &  I  & 7624           &  13 08 16.09  &  29 19 17.6   & 19.44 & 19.06 & 1.738 &  Q  \\ 
 48 &  I  & 7822           &  13 07 42.64  &  29 19 36.5   & 22.21 & 22.01 & 2.458 &  Q  \\ 
 49 &  I  & 7996           &  13 08 32.14  &  29 19 52.1   & 22.61 & 21.83 &       &     \\ 
 51 &  I  & 8169           &  13 08 01.68  &  29 20 09.1   & 21.96 & 21.47 & 0.737 &  Q  \\ 
 52 &  I  & 8184           &  13 08 22.09  &  29 20 10.5   & 23.86 & 22.63 &       &     \\ 
 53 &  I  & 8222           &  13 07 41.37  &  29 20 14.9   & 23.81 & 22.48 &       &     \\ 
 54 &  M  & 8212           &  13 07 45.40  &  29 20 13.8   & 23.46 & 22.95 &       &     \\ 
 56 &  M  & 8659           &  13 08 47.08  &  29 20 50.3   & 22.16 & 21.55 &       &     \\ 
 57 &  M  & 8754           &  13 09 01.20  &  29 20 57.4   & 27.34 & 23.15 &       &     \\ 
 60 &  I  & 8845           &  13 07 58.75  &  29 21 06.2   & 21.99 & 20.85 &       &     \\ 
 61 &  I  & 8890           &  13 08 55.51  &  29 21 10.5   & 20.96 & 19.05 &       &     \\ 
 62 &  M  & 8965           &  13 08 39.81  &  29 21 17.8   & 22.92 & 22.72 &       &     \\ 
 66 &  A  & 9215           &  13 08 56.26  &  29 21 40.6   & 21.71 & 20.41 &       &     \\ 
    &     & 9342           &  13 08 56.78  &  29 21 50.5   & 20.89 & 19.14 &       &     \\ 
 67 &  M  & 9158           &  13 07 46.95  &  29 21 34.3   & 23.02 & 22.56 &       &     \\ 
 68 &  A  & 9382           &  13 08 57.63  &  29 21 54.1   & 23.27 & 21.67 &       &     \\ 
    &     & 9418           &  13 08 57.72  &  29 21 57.4   & 22.22 & 20.44 &       &     \\ 
 69 &  I  & 9494           &  13 09 16.19  &  29 22 03.7   & 16.26 & 15.60 & 0.021 &  G  \\ 
 71 &  M  & 9820           &  13 08 56.74  &  29 22 29.0   & 21.87 & 20.10 &       &     \\ 
 73 &  I  & 9877           &  13 08 23.34  &  29 22 34.2   & 22.06 & 21.50 & 2.12  &  Q  \\ 
 76 &  I  & 9934           &  13 08 06.16  &  29 22 39.0   & 22.27 & 21.82 & 2.53  &  Q  \\ 
 77 &  I  & 9980           &  13 08 13.24  &  29 22 44.0   & 20.24 & 20.05 & 1.545 &  Q  \\ 
 79 &  I  & G5984$^{\rm d}$ &  13 08 37.00  &  29 22 48.0   & 21.07 &       &       &  G  \\ 
 80 &  I  & 10063          &  13 09 07.96  &  29 22 52.4   & 23.29 & 22.95 &       &     \\ 
 81 &  I  & 10144          &  13 09 00.18  &  29 22 58.9   & 20.18 & 18.47 &       &     \\ 
 82 &  I  & 10195          &  13 07 33.46  &  29 23 03.8   & 20.97 & 19.72 & 0.243 & BTK$^e$ \\ 
 83 &  I  & 10304          &  13 08 51.82  &  29 23 14.7   & 26.21 & 23.39 &       &     \\ 
 86 &  I  & G55866$^{\rm d}$&  13 08 19.05  &  29 23 34.8   & 23.93 &       &       &  G  \\ 
 88 &  I  & 10756          &  13 08 16.25  &  29 23 56.4   & 20.57 & 19.74 & 0.218 &  G  \\ 
 89 &  A  & 10867          &  13 07 57.48  &  29 24 06.4   & 23.24 & 23.04 &       &     \\ 
    &     & 10951          &  13 07 57.25  &  29 24 14.1   & 23.10 & 22.24 &       &     \\   
 90 &  A  & 10973          &  13 08 50.07  &  29 24 16.2   & 22.86 & 21.70 &       &     \\ 
    &     & 10953          &  13 08 50.34  &  29 24 15.0   & 20.96 & 20.07 &       &     \\   
 91 &  I  & 11014          &  13 08 27.80  &  29 24 20.7   & 17.96 & 16.57 & 0.125 &  G  \\ 
 92 &  I  & 11168          &  13 08 12.62  &  29 24 34.3   & 23.65 & 21.98 &       &     \\ 
 94 &  I  & 11320          &  13 08 47.60  &  29 24 50.8   & 24.11 & 23.50 &       &     \\ 
 95 &  I  & 11450          &  13 08 30.42  &  29 25 01.7   & 22.31 & 22.07 & 0.959 &  Q  \\ 
 96 &  I  & 11610          &  13 08 11.92  &  29 25 12.5   & 19.56 & 19.01 & 3.016 &  Q  \\ 
 97 &  I  & 11710          &  13 08 25.43  &  29 25 20.9   & 23.80 & 24.67 &       &     \\ 
 99 &  I  & 12053          &  13 07 49.24  &  29 25 48.1   & 19.64 & 18.28 & 0.241 &  G  \\ 
100 &  I  & G48612$^{\rm d}$&  13 07 41.85  &  29 26 22.9   & 22.19 & 21.85 &       &  G  \\ 
101 &  I  & G48617$^{\rm d}$&  13 08 09.04  &  29 26 25.9   & 22.42 & 22.31 &       &  G  \\ 
102 &  I  & 12472          &  13 07 29.20  &  29 26 25.4   & 22.58 & 20.93 &       &     \\ 
103 &  I  & 12467          &  13 08 34.36  &  29 26 28.2   & 20.19 & 19.69 & 0.201 &  G  \\ 
104 &  M  & 12734          &  13 08 15.54  &  29 26 50.3   & 23.79 & 23.36 &       &     \\ 
105 &  I  & 12758          &  13 08 10.16  &  29 26 52.2   & 21.99 & 20.83 &       &     \\ 
106 &  I  & 12787          &  13 08 36.58  &  29 26 56.0   & 24.20 & 22.66 &       &     \\ 
107 &  A  & 13035          &  13 09 23.96  &  29 27 15.9   & 22.94 & 20.95 &       &     \\ 
    &     & 13088          &  13 09 23.71  &  29 27 20.2   & 23.40 & 21.47 &       &     \\ 
108 &  I  & 13155          &  13 08 08.73  &  29 27 26.8   & 21.71 & 21.06 &       & [Q]$^e$ \\ 
109 &  M  & 13310          &  13 07 56.71  &  29 27 38.0   & 21.66 & 21.07 &       &     \\ 
110 &  I  & 13360          &  13 08 13.16  &  29 27 43.5   & 22.86 & 22.81 &       &     \\ 
111 &  I  & 13412          &  13 09 06.34  &  29 27 48.1   & 22.11 & 21.85 & 2.082 &  Q  \\ 
113 &  M  & G55532$^{\rm d}$&  13 08 53.56  &  29 27 51.1   & 23.81 &       &       &  G  \\ 
114 &  I  & 13459          &  13 07 51.79  &  29 27 51.6   & 22.83 & 22.00 &       & BTK$^e$ \\ 
115 &  I  & 13732          &  13 08 24.04  &  29 28 19.2   & 22.76 & 22.64 &       &     \\ 
116 &  I  & 13966          &  13 08 44.15  &  29 28 40.3   & 21.30 & 20.96 & 0.951 &  Q  \\ 
117 &  I  & 13993          &  13 08 44.83  &  29 28 43.6   & 23.16 & 21.71 &       &     \\ 
118 &  A  & 14002          &  13 08 55.19  &  29 28 45.0   & 24.23 & 22.03 &       &     \\ 
    &     & 14128          &  13 08 55.44  &  29 28 57.1   & 21.96 & 20.19 &       &     \\   
120 &  I  & 14264          &  13 08 03.40  &  29 29 08.8   & 20.39 & 18.78 & 0.288 & BTK$^e$ \\ 
122 &  I  & 14630          &  13 08 51.24  &  29 29 45.4   & 23.82 & 22.93 &       &     \\ 
123 &  A  & 14733          &  13 08 20.29  &  29 29 56.3   & 23.21 & 21.93 &       &     \\ 
    &     & 14728          &  13 08 20.37  &  29 29 56.6   & 23.19 & 21.64 &       &     \\ 
    &     & 14795          &  13 08 20.55  &  29 30 01.1   & 20.77 & 19.01 & 0.318 &  G  \\   
    &     & G59730$^{\rm d}$&  13 08 20.02  &  29 29 58.1   & 22.98 &       &       &  G  \\ 
    &     & G59724$^{\rm d}$&  13 08 20.33  &  29 30 02.9   & 20.19 &       &       &  G  \\ 
124 &  A  & 15102          &  13 08 12.16  &  29 30 27.5   & 23.67 & 23.24 &       &     \\ 
    &     & 15093          &  13 08 11.92  &  29 30 27.7   & 23.38 & 23.29 &       &     \\   
125 &  I  & 15180          &  13 08 24.40  &  29 30 36.9   & 21.13 & 20.71 & 1.797 &  Q  \\ 
126 &  I  & 15248          &  13 08 42.15  &  29 30 43.9   & 22.28 & 21.79 & 1.145 &  Q$^e$  \\ 
127 &  I  & 15465          &  13 09 17.09  &  29 31 04.3   & 21.89 & 21.15 &       & [Q]$^e$ \\ 
131 &  I  & 16251          &  13 08 31.73  &  29 32 20.9   & 23.68 & 22.21 &       &     \\ 
134 &  I  & 16677          &  13 08 49.85  &  29 33 03.3   & 23.45 & 23.25 &       &     \\ 
135 &  M  & 16713          &  13 07 49.36  &  29 33 03.3   & 19.73 & 19.20 & 0.993 &  Q  \\ 
136 &  I  & 17001          &  13 08 06.24  &  29 33 32.4   & 23.17 & 23.19 &       & BTK$^e$ \\ 
137 &  A  & 17231          &  13 08 47.04  &  29 33 56.1   & 24.33 &       &       &     \\ 
    &     & 17174          &  13 08 47.39  &  29 33 50.5   & 23.64 & 22.14 &       &     \\ 
139 &  I  & 17494          &  13 08 33.43  &  29 34 23.2   & 23.37 & 22.51 & 0.75  &  RG \\ 
140 &  I  & 17750          &  13 08 08.51  &  29 34 49.1   & 19.79 & 19.23 & 1.179 &  Q  \\ 
\end{longtable}      
\noindent
$^{\rm a}$ Rank of optical identification: I=identified, M=marginal identification, A=ambiguous identification.\\
$^{\rm b}$ Serial number in the optical catalogue of the KPNO survey of SA 57 \citep{kron80,koo86}.\\
$^{\rm c}$ Q: quasar; G: galaxy; BTK extended variable object from \citet{btk98}; RG: radio galaxy; [Q]: point-like quasar candidate.\\
$^{\rm d}$ Serial number in the North Galactic Pole Faint Galaxy Catalogue \citep{infa95}.\\
$^{\rm e}$ Candidates or confirmed AGNs selected through variability and not preaviously selected by other methods.\\

\begin{longtable}{rccccc}
\caption{\label{Tab4} X-ray undetected optical sources}\\
\hline\hline
NSER  &RA(2000)&DEC(2000)& $F$ &  $z$  & $f_x(2-10{\rm ~keV})$                 \\ 
      &  &   &       &       & [$10^{-14}{\rm ~erg/cm^2/s}$] \\   
\hline
\endfirsthead
\endhead
\hline
\endfoot
 5326 & 13 07 32.5 & 29 15 45 & 20.066 & 2.86 & $<$4.16 \\
 5482 & 13 09 08.8 & 29 15 58 & 20.747 & 0.45 & $<$1.10 \\
 5748 & 13 09 23.6 & 29 16 20 & 19.890 & 0.21 & $<$2.04 \\
 7042 & 13 07 26.2 & 29 18 25 & 21.590 & 0.53 & $<$3.03 \\
 7701 & 13 07 35.1 & 29 19 24 & 22.073 & 0.99 & $<$0.75 \\
 8119 & 13 08 44.2 & 29 20 05 & 18.103 & 0.745 & $<$0.99 \\
 8303 & 13 09 12.8 & 29 20 20 & 18.112 & 0.72 & $<$0.91 \\
 8945 & 13 08 37.8 & 29 21 15 & 21.733 & 0.61 & $<$0.43 \\
 9542 & 13 07 42.4 & 29 22 06 & 19.299 & 0.233 & $<$0.87 \\
10601 & 13 08 47.8 & 29 23 41 & 21.050 & 0.44 & $<$0.57 \\
11255 & 13 07 28.4 & 29 24 41 & 21.971 & 0.18 & $<$2.91 \\
11334 & 13 07 52.1 & 29 24 50 & 21.806 & 1.82 & $<$0.48 \\
11335 & 13 07 22.0 & 29 24 48 & 18.295 & 3.36 & $<$3.44 \\
12536 & 13 09 06.8 & 29 26 32 & 21.403 & 0.42 & $<$1.52 \\
12573 & 13 07 48.9 & 29 26 34 & 17.583 & 2.20 & $<$1.31 \\
14054 & 13 07 59.3 & 29 28 49 & 18.373 & 0.1357 & $<$1.05 \\
15041 & 13 08 03.6 & 29 30 20 & 18.331 & 0.137 & $<$1.12 \\
15437 & 13 09 12.7 & 29 31 01 & 18.078 & 2.82 & $<$2.59 \\
16626 & 13 08 13.4 & 29 32 55 & 17.820 & 2.95 & $<$0.58 \\
\hline               
\end{longtable}      

\end{document}